\documentclass{INTERSPEECH2023}

 \interspeechcameraready

\usepackage[utf8]{inputenc}
\usepackage{algorithm}
\usepackage{algorithmic}
\usepackage{cite}
\usepackage{bm}
\usepackage{hyperref}
\usepackage{comment}
\usepackage{pifont}
\newcommand{\cmark}{\ding{51}}%
\usepackage{subcaption}
\usepackage{xcolor}

\title{4D ASR: Joint modeling of CTC, Attention, Transducer, \\and Mask-Predict decoders}

\name{Yui Sudo$^1$, Muhammad Shakeel$^1$, Brian Yan$^2$, Jiatong Shi$^2$, Shinji Watanabe$^2$}
\address{
  $^1$Honda Research Institute Japan Co., Ltd., Saitama, Japan\\
  $^2$Language Technologies Institute, Carnegie Mellon University, Pittsburgh, PA, USA}
\email{\{yui.sudo, shakeel.muhammad\}@jp.honda-ri.com, \\ \{byan, jiatongs\}@andrew.cmu.edu, shinjiw@ieee.org}

\begin{document}
%\ninept
%
\maketitle
\begin{abstract}

End-to-end (E2E) automatic speech recognition (ASR) can be classified into several models, including connectionist temporal classification (CTC), recurrent neural network transducer (RNN-T), attention mechanism, and mask-predict models.
There are pros and cons to each of these architectures, and thus practitioners may switch between these different models depending on application requirements.
Instead of building separate models, we propose a joint modeling scheme where four different decoders (CTC, attention, RNN-T, mask-predict) share an encoder -- we refer to this as 4D modeling.
Additionally, we propose to 1) train 4D models using a two-stage strategy which stabilizes multitask learning and 2) decode 4D models using a novel time-synchronous one-pass beam search.
We demonstrate that jointly trained 4D models improve the performances of each individual decoder. Further, we show that our joint CTC/RNN-T/attention decoding surpasses the previously proposed CTC/attention decoding.
\end{abstract}


\section{Introduction}
\label{sec:intro}

End-to-end (E2E) automatic speech recognition (ASR) has been actively studied. E2E ASR systems include four main network architectures, such as connectionist temporal classification (CTC) \cite{ctc1,ctc2,kriman2020quartznet}, recurrent neural network transducer (RNN-T) \cite{rnnt1,zhang2020transformer,han2020contextnet,rnnt2,lee2021intermediate}, attention mechanism \cite{attention1,attention2,karita2019comparative,guo2021recent}, and non-autoregressive (NAR) methods \cite{higuchi2020mask,chen2020non,song2021non}. 
These networks align speech signals and token sequences in various ways, each with its own strengths and weaknesses, as follows:
\begin{itemize}
%\vspace*{-2mm}
%\leftskip -3.5mm 
%\itemsep -0.5mm
    \item 
    %Under the assumption of conditional independence, %(i.e., each frame-level prediction can be generated independently), 
    %CTC predicts a monotonic alignment of output tokens with input frames of speech. The CTC's high speed makes it suitable for applications that require real-time performance. However, it may perform poorly due to the conditional independent assumption. 
    CTC assumes conditional independence and predicts monotonic alignment of output tokens with input frames. It is fast and suitable for real-time applications, but its performance may suffer from the conditional independence assumption.
    CTC can also be used to segment long recordings \cite{ctcsegmentation}.
    \item RNN-T also has a monotonic alignment assumption, but unlike CTC, it relaxes the conditional independence assumption. It generally outperforms CTC and is suitable for streaming ASR \cite{rnnt2}. However, the modeling space is larger than that of the CTC, making it more difficult to train.
    \item The attention model includes a source-target attention mechanism that aligns speech signals with token sequences. This mechanism is extremely useful in tasks requiring flexible alignments between input and output sequences (e.g., translation tasks) \cite{luong2015effective,sperber2019attention}. However, it is prone to alignment errors in ASR tasks because it lacks the monotonicity constraint.
    \item A typical example of NAR models is Mask-CTC \cite{higuchi2020mask}. Mask-CTC estimates the token sequence using the entire sequence. The mask-based approach \cite{ghazvininejad2019mask} can consider label dependency and still retain fast latency. It can be also used for two-pass rescoring approaches as ASR error correction \cite{futami2022correction,wang2022deliberation}.
\vspace*{-4mm}
\end{itemize}

\begin{figure}[t!]
\vspace{-4mm}
    \centering
        \begin{minipage}{0.36\textwidth}
            \includegraphics[width=\textwidth]{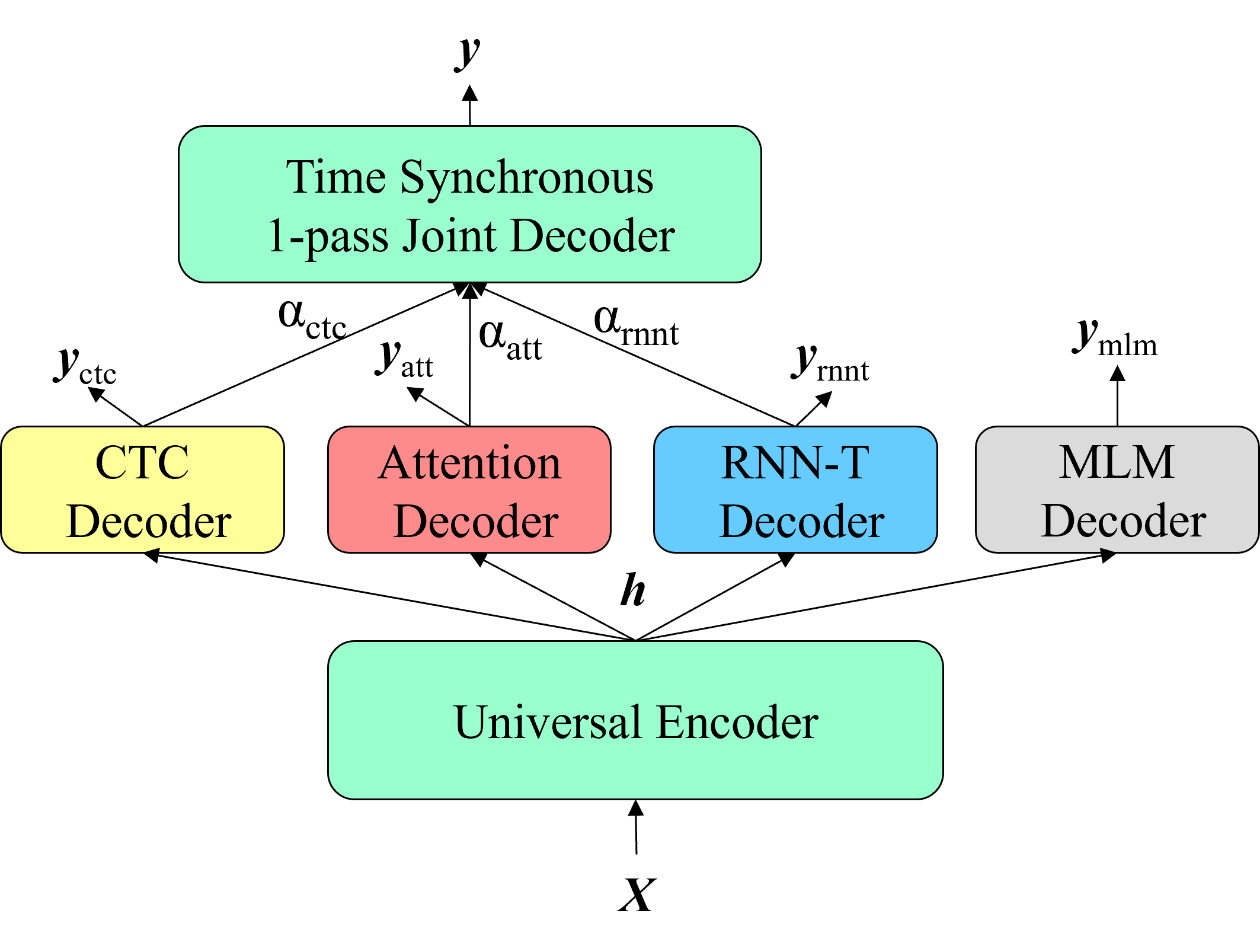} 
        \end{minipage}
        \centering
    \vspace*{-5mm}
    \caption{4D joint model of CTC, attention, RNN-T, and Mask-CTC: the four decoders share a universal encoder and joint decoding is performed using CTC, attention, and RNN-T.} 
    \label{fig:mtl}
\vspace*{-5mm}
\end{figure}

Since each network architecture has the above different properties, separate models for each application scenario are usually required. CTC, for example, is appropriate for on-device systems requiring less computation, whereas attention is suitable for offline systems with less stringent latency requirements. However, having multiple models for each application scenario increases overheads.
Several attempts have been made to integrate these models, such as CTC/attention, to reduce their respective shortcomings \cite{watanabe2017hybrid,Kim,ueno2018acoustic}. %,nakatani2019improving}. 
Multitask learning is performed by sharing an encoder to regularize the over-flexibility of the attention mechanism and the conditional independence assumption of the CTC model. 
A one-pass beam search is performed during decoding using both CTC and attention decoders to further improve performance \cite{hori2017joint}. 
Other integrated models based on two-pass decoding of RNN-T/attention, NAR/attention, and NAR/RNN-T models have also been proposed \cite{sainath2019two,hu2021transformer,tian2022hybrid,wang2022deliberation,yao2021wenet}.
Given the success of these joint models with two types of decoders, a natural question is \textit{how many decoders can be successfully integrated in a single model?}

In this work, we seek to jointly model four decoders (4D) with a shared encoder (Figure \ref{fig:mtl}): CTC, attention, RNN-T, and Mask-CTC.
To accommodate the increased modeling complexity under this 4D scheme, we adapt both our training and decoding strategies.
In particular, we:
\begin{itemize}
    \item Employ a two-stage optimization strategy to select multitasking hyper-parameters in an efficient, data-driven manner.
    \item Demonstrate experimentally that each single-decoder branch of 4D models are improved over counterparts without joint training across three benchmark ASR tasks.
    \item Further introduce novel time-synchronous beam search algorithms for joint CTC/RNN-T/attention decoding, which outperforms CTC/attention decoding on average.
\end{itemize}

\section{Preliminary}
\label{sec:Preliminary}

%\vspace*{-1mm}
This section describes our Conformer-based encoder, which is commonly used in the four models, CTC, attention, RNN-T, and Mask-CTC. Then, we describe each decoder in detail.

\vspace*{-0mm}
\subsection{Universal encoder}
\label{sec:encoder}
%\vspace*{-1mm}
Conformer \cite{guo2021recent,gulati2020conformer} is used as a universal encoder in our study, which consists of two convolutional layers, a linear projection layer, and a positional encoding layer, followed by Conformer blocks. 
The convolutional layers subsample an audio feature sequence, \begin{math}\bm{X}\end{math}, into a subsampled feature sequence. %, \begin{math}\bm{u}\end{math}. 
Then, the Conformer blocks transform the subsampled feature sequence to a \begin{math}T\end{math}-length hidden state sequence,
\begin{math}
\bm{H} = [\bm{h}_1, ... , \bm{h}_T], 
\end{math}
described as,
\begin{equation}
\label{transformer-encoder}
\bm{H} = \mathrm{ConEncoder}(\bm{X}).
\end{equation}
%Each Conformer block has a multiheaded self-attention layer, a linear layer, and a layer-normalization, with residual connections.% \cite{ba2016layer,7780459}. 

\vspace*{-0mm}
\subsection{CTC}
\label{sec:ctc}

%Given \begin{math}\bm{h}\end{math} generated by the encoder in Eq. (\ref{transformer-encoder}), 
The CTC decoder estimates the output token sequence, \begin{math}\bm{y}\end{math},
given \begin{math}\bm{H}\end{math} generated by the encoder in Eq. \eqref{transformer-encoder}.
%\begin{math}
%\bm{y} = \mathrm{CTCDecoder}(\bm{h}).
%\end{math}
CTC introduces the alignment sequence \begin{math}\bm{z} = \{\bm{z}_t \in V \cup \{\phi\}\}\end{math}, where $t$, $V$, and \begin{math}\phi\end{math} denote the time index, the vocabulary, and a blank token, respectively.
%denotes the time index, %(\begin{math}1 \leq t \leq T\end{math}) 
%$V$ denotes the vocabulary, and \begin{math}\phi\end{math} denotes a blank token. 
%Instead of estimating output posteriors, \begin{math}P(\bm{y}|\bm{X})\end{math}, 
CTC estimates the alignment posteriors, \begin{math}P_{\text{}}(\bm{z}|\bm{X})\end{math}. Each alignment sequence, \begin{math}\bm{z}\end{math}, is deterministically mapped to a corresponding $S$-length output sequence, \begin{math}\bm{y} = [y_1, ... ,y_S]\end{math}.
CTC assumes conditional independence, yielding: %estimate of Eq. \eqref{eq:ctc}:
\begin{equation}
\label{eq:ctc}
P_{\text{ctc}}(\bm{y} \mid \bm{X}) = \sum_{\bm{z} \in \mathcal{Z_{\text{}}}(\bm{y})} P_{\text{}}(\bm{z} \mid \bm{X}) = \sum_{\bm{z} \in \mathcal{Z_{\text{}}}(\bm{y})} \left[\prod_{t=1}^{T} P\left(\bm{z}_{t} \mid \bm{h}_t\right)\right].
%P_{\text{ctc}}(\bm{z} \mid \bm{h}) = \prod_{t=1}^{T} P\left(z_{t} \mid h_t\right),
\end{equation}
During training, CTC optimizes model parameters by minimizing the following negative log-likelihood as follows:
\begin{equation}
%L_{\text{ctc}} = - \log \sum _{\bm{z} \in \mathcal{Z}(\bm{y})} P_{\text{ctc}}(\bm{z} \mid \bm{h}),
L_{\text{ctc}} = - \log P_{\text{ctc}}(\bm{y} \mid \bm{X}),
\label{eq:loss_ctc}
\end{equation}
where $\mathcal{Z}(\bm{y})$ is a set of all possible alignment sequences of~$\bm{y}$.

\vspace*{-0mm}
\subsection{RNN-T}
\label{sec:rnnt}

The RNN-T decoder comprises a prediction network and a joint network. 
The prediction network generates a high-level representation \begin{math}g_s\end{math} by conditioning on the previous non-blank token sequence \begin{math}\bm{g}_{s-1}\end{math}, where \begin{math}s\end{math} denotes a non-blank token index. %, unlike CTC in Eq. \eqref{eq:ctc}. % as
%\begin{math}
%l_s = \mathrm{PredNet}(\bm{l_{s-1}}).
%\end{math}
The joint network is a feed-forward network that combines $\bm{h}_t$ and  $g_{s}$. % as
%\begin{math}
%z_{t,s} = \mathrm{JointNet}(h_t, g_{s}).
%\end{math}
While CTC assumes the conditional independence in Eq.~\eqref{eq:ctc}, 
the RNN-T model marginalizes the potential alignments $\bm{u}$ that output \begin{math}\bm{y}\end{math} as follows:
\begin{equation}
\label{rnntlikelihood}
P_{\text{rnnt}}(\bm{y} \mid \bm{X}) = \sum_{\bm{z} \in \mathcal{Z}(\bm{y})} P(\bm{z} | \bm{X}) = \sum_{\bm{z} \in \mathcal{Z}(\bm{y})} \left[\prod_{i=1}^{T+S} P\left(\bm{z}_{i} \mid \bm{h}_{t_i}, g_{s_i}\right)\right],
\end{equation}
where $i$ represents a position in $(T+S)$-length alignment path specified by $t_i$-th decoder state and $s_i$-th token, respectively. %$S$ denotes the total length of the complete token sequence, \begin{math}\bm{y}\end{math}.
%where \begin{math}\bm{h}_{t_i}\end{math} denotes hidden state at $t$ and \begin{math}\bm{g}_{s_i}\end{math} denotes $s$-th token, and $i$ denotes the index that represents the sum of $t$ and $s$.
RNN-T optimizes model parameters by minimizing the following negative log-likelihood described as, 
\begin{equation}
L_{\text{rnnt}} = - \log P_{\text{rnnt}}(\bm{y} \mid \bm{X}).
%L_{\text{rnnt}} = - \log \sum_{\bm{z} \in \mathcal{Z}(\bm{y})} p(\bm{z} | \bm{X}).
\label{eq:loss_rnnt}
\end{equation}

\vspace*{-0mm}
\subsection{Attention}
\label{sec:attention}

Given \begin{math}\bm{H}\end{math} generated by the encoder in Eq. \eqref{transformer-encoder} 
and the previously estimated token sequence \begin{math}\bm{y}_{s-1}\end{math}, the attention decoder recursively estimates the next token \begin{math}y_s\end{math}. %, where \begin{math}s\end{math} denotes a token index. 
The token history $\bm{y}_{s-1}$ is converted to token embeddings and fed into decoder layers with hidden states $\bm{H}$, unlike $\bm{h}_t$ in CTC/RNN-T. %The decoder then predicts the probability of $y_s$.
While CTC and RNN-T uses $\bm{h}_t$ in Eq. \eqref{eq:ctc} and (\ref{rnntlikelihood}), the likelihood of an attention model is described as follows:
%\vspace*{-2mm}
\begin{equation}
\label{attlikelihood}
P_{\text{att}}(\bm{y} \mid \bm{X})=\prod_{s=1}^{S} P\left(y_{s} \mid \bm{y_{s-1}}, \bm{H}\right).
\end{equation}
Attention optimizes model parameters by minimizing the following negative log-likelihood described as, 
\begin{equation}
L_{\text{att}} = - \log P_{\text{att}}(\bm{y} \mid \bm{X}).
\label{eq:loss_att}
\end{equation}

\vspace*{-0mm}
\subsection{Mask-CTC}
\label{sec:maskctc}

The Mask-CTC model uses a masked language model (MLM) \cite{ghazvininejad2019mask} decoder, which estimates the token sequence using the entire sequence given \begin{math}\bm{H}\end{math} in Eq. \eqref{transformer-encoder}, smilar to the attention case. %described as,
%\begin{math}
%\bm{y} = \mathrm{MLMDecoder}(\bm{h}).
%\end{math}
However, unlike attention, randomly sampled tokens \begin{math}y_{\text{mask}}\end{math} are masked with a special token during training. Then, \begin{math}y_{\text{mask}}\end{math} is predicted conditioning on the remaining unmasked tokens \begin{math}y_{\text{obs}}\end{math} as %,
\begin{math}
P_{\text{mlm}}(y_{\text{mask}} | y_{\text{obs}}, \bm{X}).
\end{math}
Mask-CTC optimizes model parameters by minimizing the following negative log-likelihood:
\begin{equation}
L_{\text{mlm}} = - \log P_{\text{mlm}}(y_{\text{mask}}|y_{\text{obs}}, \bm{X}).
\label{eq:loss_mlm}
\end{equation}

\vspace*{0mm}
\section{Proposed 4D ASR model}
\label{sec:proposed}
\vspace*{0mm}

This section describes the joint training using the two-stage optimization strategy and the joint decoding used in the proposed 4D model as shown in Figure \ref{fig:mtl}.
%Figure \ref{fig:mtl} shows the overall structure of the proposed method. The four decoders share a single encoder and joint decoding is performed using CTC, attention, and RNN-T. This section describes the joint training and one-pass joint decoding.

\vspace*{-0mm}
\subsection{Joint training with a two stage strategy}
\label{sec:jointraining}

Multitask learning is performed using the weighted sum of losses shown in Eqs.~\eqref{eq:loss_ctc}, \eqref{eq:loss_rnnt}, \eqref{eq:loss_att}, and \eqref{eq:loss_mlm} described as follows:
\vspace*{1mm}
\begin{equation}
L = \lambda_{\text{ctc}} L_{\text{ctc}} + \lambda_{\text{rnnt}} L_{\text{rnnt}} + \lambda_{\text{att}} L_{\text{att}} + \lambda_{\text{mlm}} L_{\text{mlm}},
\label{eq:multitask}
\vspace*{1mm}
\end{equation}
where $\lambda$ represents training weights.
Training weights are usually determined experimentally \cite{Kim} or based on meta-learning \cite{Lin2019AdaptiveAT}. 
%Franceschi2017ForwardAR doublecheck later
In this work, with four weights, experimenting with all possible combinations would be overly time-consuming. 
To address this issue, we used a two-stage optimization strategy to determine the multitask weights $(\lambda_{\text{ctc}}, \lambda_{\text{rnnt}}, \lambda_{\text{att}}, \lambda_{\text{mlm}})$ in Eq.~\eqref{eq:multitask}. 
In the first stage, all four training weights were set to be equal, i.e., $(0.25, 0.25, 0.25, 0.25)$. 
%All four training weights were set to be equal in the first stage $(0.25, 0.25, 0.25, 0.25)$. 
Then, in the second stage, the training weights were determined to be roughly proportional to the number of epochs in the first stage at which each validation loss reached its minimum value. 
%The training weights in the second stage were then determined to be roughly proportional to the number of epochs in the first stage, at which each validation loss showed the minimum value. 
For example, if the validation losses ($L_{\text{ctc}}, L_{\text{rnnt}}, L_{\text{att}}, L_{\text{mlm}}$) takes their minimum values at the 10th, 10th, 10th, and 70th epochs in the first stage, the training weights in the second stage were set to $(0.1, 0.1, 0.1, 0.7)$. 
This strategy is based on the proposition that losses requiring more epochs to convergence should be given higher weights.

\begin{figure*}[t!]
%\vspace{-6mm}
    \centering
        \begin{minipage}{1.0\textwidth}
            \includegraphics[width=0.99\textwidth]{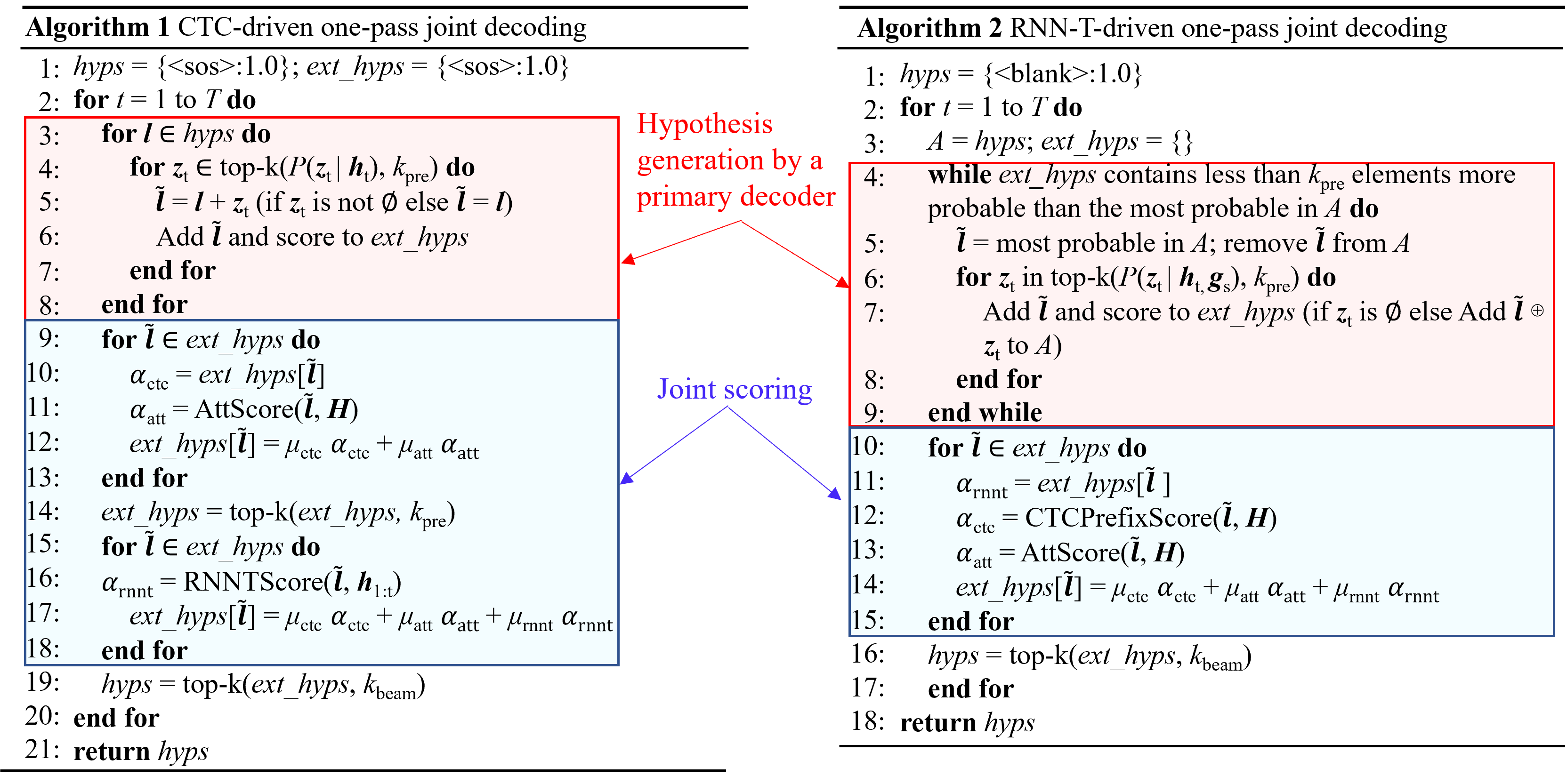} 
        \end{minipage}
        \centering
    \label{fig:algorithm}
\vspace*{-6mm}
\end{figure*}

\vspace*{-0mm}
\subsection{Time synchronous one-pass joint decoding}
\label{sec:joint docoder}
%\vspace*{-2mm}
Another key contribution of this paper is to propose two one-pass joint CTC/RNN-T/attention decoding algorithms using time-synchrony: Algorithms 1 and 2 present CTC-driven and RNN-T-driven algorithms, respectively. 
While adding an RNN-T decoder to label-synchronous one-pass beam search proposed in \cite{watanabe2017hybrid} would require computing all RNN-T alignment paths, the proposed method avoids this inefficient computation by using time synchrony.
Note that we excluded the Mask-CTC from the proposed joint decoding because, unlike the other three decoders, it predicts the entire output sequence in parallel. 

The CTC-driven method uses CTC as a primary decoder, which generates initial hypotheses, \textit{ext\_hyps} (lines 3-8 in Algorithm 1); this is similar to CTC-driven joint decoding methods with attention \cite{moritz2019triggered, yan2022ctc} or a language model \cite{hannun2014first}, but additionally accounts for RNN-T likelihoods.
The RNN-T-driven method uses RNN-T as a primary (lines 4-9 in Algorithm 2); this time-synchronous method is performed autoregressively, unlike iterative CTC refinement of RNN-T \cite{wang2022deliberation}, plus we additionally account for attention likelihoods.

Then, the generated hypotheses, \textit{ext\_hyps}, are scored combining CTC, attention, and RNN-T decoders (lines 9-18 in Algorithm 1, lines 10-15 in Algorithm 2).
The attention score is calculated using the forward computation as in Eq.~(\ref{attlikelihood}) for both joint decoding methods.
The CTC score is calculated using dynamic programming as in \cite{ctc1} for the CTC-driven method, and CTC prefix scoring proposed in \cite{watanabe2017hybrid} is used for the RNN-T-driven method.
As for RNN-T scoring, the probabilities of all possible paths from the previous hypotheses at $t-1$ to the current hypothesis at $t$ are added for the CTC-driven method, whereas the RNN-T score is calculated as in the conventional RNN-T \cite{rnnt1}.
%Each decoder score $\alpha$, and length penalty $\beta$ \cite{watanabe2017hybrid} are added with the 
Each decoder score $\alpha$ is added with the decoder weights ($\mu_{\text{ctc}}$, $\mu_{\text{att}}$, $\mu_{\text{rnnt}}$) (line 17 in Algorithm 1, line~14 in Algorithm 2). 
%\begin{equation}
%\textit{ext\_hyps}[l] = \alpha_{\text{joint}} = \mu_{\text{ctc}} \alpha_{\text{ctc}} +  %\mu_{\text{att}} \alpha_{\text{att}} + \mu_{\text{rnnt}} \alpha_{\text{rnnt}}.
%\label{eq:weight}
%\end{equation}

Top $k_\text{beam}$ hypotheses, \textit{hyps}, are retained for the next time frame based on the obtained joint score (line 19 in Algorithm 1, line 16 in Algorithm 2), where $k_\text{beam}$ denotes the main beam size.

Note that, RNN-T-driven joint decoding generates only $k_\text{pre}$ hypotheses (line 4 in Algorithm 2), whereas CTC-driven joint decoding generates $k_\text{pre} \times k_\text{beam}$ hypotheses (line 3-8 in Algorithm 1). In other words, CTC-driven joint decoding requires roughly $k_\text{beam}$ times more computation for scoring.
Therefore, we pruned the hypotheses in CTC-driven joint decoding to reduce the computational cost. The top $k_\text{pre}$ hypotheses were chosen specifically based on the CTC and attention %, and length penalty 
scores before RNN-T scoring (line 14 in Algorithm 1).

\begin{table*}[t]
\caption{Our 4D models with jointly trained encoders (\texttt{B1-8}) compared to respective baselines (\texttt{A1-5}). Best WER/CER result in each comparison is \textbf{bolded} and best results overall are further \underline{\textbf{underlined}}. The average absolute improvements ($\Delta$) are also shown.}
\vspace*{-11mm}
\label{maintable}
\begin{center}

\resizebox {0.96\linewidth} {!} {
% \begin{tabular}{@{}cc|cc|cc|ccc}
% \hline
%             &           & \multicolumn{2}{c|}{LibriSpeech 960 h} & \multicolumn{2}{c|}{LibriSpeech 100 h} & \multicolumn{3}{c}{CSJ (APS subset)} \\
% Training     & Decoder   & test-clean & test-other & test-clean & test-other & eval1 & eval2 & eval3 \\
% \hline
% & Mask-CTC  & 2.97 & 6.92 & 8.72 & 20.77 & 5.54 & 4.01 & 9.51 \\
%  & CTC       & 3.10 & 6.90 & 8.20 & 20.66 & 5.50 & 3.88 & 9.75 \\
% Baseline    & Attention  & 2.88 & 5.62 & 8.59 & 19.43 & 5.14 & 3.86 & 9.47 \\
%  & RNN-T & 2.66 & 5.82 & 7.25 & 18.30 & 5.60 & 4.09 & 9.90\\
%     & CTC/Attention  & 2.45 & \textbf{5.18} & 7.08 & 17.80 & \textbf{5.10} & \textbf{3.61} & 9.17\\
% \hline
%     & Attention  & 2.74 & 5.65 & 7.83 & 17.91 & 5.12 & 3.79 & 9.25\\
%     & CTC/Attention  & 2.42 & 5.31 & 6.49 & 17.03 & 5.11 & 3.72 & 8.80\\
% 4D multitasking & RNN-T/Attention & \textbf{2.37} & 5.25 & 6.35 & 16.47& 5.13 & 3.73 & 8.98\\
%    & CTC-driven CTC/RNN-T/Att & 2.42 & 5.25 & \textbf{6.32} & 16.48 & 5.15 & 3.68 & \textbf{8.75}\\
%     & RNN-T-driven CTC/RNN-T/Att & 2.38 & 5.21 & 6.33 & \textbf{16.43}& 5.14 & 3.64 & 8.88\\
% \hline% 
\begin{tabular}{clc|cc|cc|cc|c}
\toprule

 & & \textsc{Joint} & \multicolumn{2}{c|}{\textsc{LibriSpeech 960 h}} & \multicolumn{2}{c|}{\textsc{LibriSpeech 100 h}} & \multicolumn{2}{c|}{\textsc{In-house}}  \\
 \cmidrule(lr){4-5}  \cmidrule(lr){6-7}  \cmidrule(lr){8-9}  
\texttt{ID} & \textsc{Model Name} & \textsc{Encoder} & test-clean & test-other & test-clean & test-other & assembly & meeting & avg $\Delta$ \\
\midrule
\texttt{A1} & Attention \textit{(baseline)} & - & 2.88 & \textbf{5.62} & 8.59 & 19.43 & 3.81 & 5.82 & - \\
\texttt{B1} & 4D Attention & \cmark & \textbf{2.66} & \textbf{5.62} & \textbf{7.83} & \textbf{17.91} & \textbf{3.56} & \textbf{5.31} & -0.54\\

\midrule
\texttt{A2} & CTC \textit{(baseline)} & & 3.10 & 6.90 & 8.20 & 20.66 & 3.91 & 6.30 & -\\
\texttt{B2} & 4D CTC & \cmark & \textbf{2.84} & \textbf{6.39} & \textbf{7.30} & \textbf{18.96} & \textbf{3.74} & \textbf{5.53} & -0.72 \\

\midrule

\texttt{A3} & Mask-CTC \textit{(baseline)} & - & \textbf{2.97} & 6.92 & 8.72 & 20.78 & 4.37 & 6.56 & -\\
\texttt{B3} & 4D Mask-CTC & \cmark & 3.11 & \textbf{6.82} & \textbf{7.47} & \textbf{18.97} & \textbf{3.86} & \textbf{6.24} & -0.64\\

\midrule

\texttt{A4} & RNN-T \textit{(baseline)} & - &  2.66 & 5.82 & 7.25 & 18.30 & 4.00 & 5.91 & - \\
\texttt{B4} & 4D RNN-T & \cmark & \textbf{2.56} & \textbf{5.74} & \textbf{7.10} & \textbf{17.61} & \textbf{3.94} & \textbf{5.30} & -0.28 \\

\midrule

\texttt{A5} & CTC/Attention \textit{(baseline)} & - & 2.45 & \textbf{5.18} & 7.08 & 17.80 & 3.75 & 5.72 & - \\
\texttt{B5} & 4D CTC/Attention & \cmark & \textbf{2.42} & 5.31 & \textbf{6.49} & \textbf{17.03} & \textbf{3.67} & \textbf{5.17} & -0.32 \\

\midrule

\texttt{A5} & CTC/Attention \textit{(baseline)} & - & 2.45 & \underline{\textbf{5.18}} & 7.08 & 17.80 & 3.75 & 5.72 & - \\

\texttt{B6} & 4D RNN-T/Attention (RNN-T-driven) & \cmark & \underline{\textbf{2.37}} & 5.25 & \textbf{6.35} & \textbf{16.47} & 3.81 & \textbf{5.17} & -0.43 \\
\texttt{B7} & 4D CTC/RNN-T/Attn (CTC-driven) & \cmark & \textbf{2.42} & 5.25 & \underline{\textbf{6.32}} & \textbf{16.48} & \textbf{3.69} & \underline{\textbf{5.12}} & -0.45 \\
\texttt{B8} & 4D CTC/RNN-T/Attn (RNN-T-driven) & \cmark & \textbf{2.38} & 5.21 & \textbf{6.33} & \underline{\textbf{16.43}} & \underline{\textbf{3.65}} & \textbf{5.16} & -0.47 \\
\bottomrule
\end{tabular}
}

\end{center}
\vspace*{-5mm}
\end{table*}

\vspace*{-0mm}
\section{Experiments}
\label{sec:experiments}

%We evaluate the proposed method in terms of performance, effectiveness of the two-stage training strategy, and computational cost during decoding.

\vspace*{-0mm}
\subsection{Experimental setup}
\label{sec:experimental condition}
\vspace*{-1mm}

The input features were 80-dimensional Mel-scale filter-bank features with a window size of 512 samples and a hop length of 160 samples. The sampling frequency was 16 kHz. SpecAugment \cite{specaug} was then used.
The encoder consisted of two convolutional layers %with stride two
and a 512-dimensional linear projection layer followed by 12 Conformer layers with 2048 linear units. % and layer normalization. 
The CTC decoder had a 1-layer linear layer. The attention and MLM decoders had six Transformer layers with 2048 linear units each. We set the attention dimension size to 512 with 8-multi-head attention. The RNN-T decoder used a 1-layer long short-term memory (LSTM) with a 512 hidden size and a linear layer of 640 joint sizes for the prediction and joint networks, respectively. 
The proposed model was trained 150 epochs using the Adam optimizer at a learning rate of 0.0015.
%\cite{kingma2014adam}
%with warmup steps of 15000. 
The training weights ($\lambda_{\text{ctc}}$, $\lambda_{\text{rnnt}}$, $\lambda_{\text{att}}$, $\lambda_{\text{mlm}}$) of the second stage were (0.15, 0.10, 0.30, 0.45) based on the two-stage strategy described in Section \ref{sec:jointraining}. %, %which will be discussed in Section \ref{sec:joint training}. 
The decoder weights ($\mu_{\text{ctc}}$, $\mu_{\text{rnnt}}$, $\mu_{\text{att}}$) of CTC-driven and RNN-T-driven joint decoding in Algorithms 1 and 2 were (0.2, 0.2, 0.6) and (0.1, 0.4, 0.5), respectively. 

%Three datasets, LibriSpeech (960 h, 100 h) \cite{panayotov2015librispeech} and our in-house dataset were used to test the proposed method. 
The proposed method was tested using the LibriSpeech (960 h, 100 h) \cite{panayotov2015librispeech} and our in-house dataset (855 h).
Our in-house dataset\footnote{Our in-house dataset is not released for privacy and confidentiality  reasons.} consists of 93 hours of Japanese speech data, % collected from different locations, 
including meeting and morning assembly scenarios,
plus the Corpus of Spontaneous Japanese (581 h)~\cite{csj} and the 181 hours of Japanese speech database developed by the Advanced Telecommunications Research Institute International (ATR-APP)~\cite{KUREMATSU1990357}.
%The word error rate (WER) and character error rate (CER) were calculated for LibriSpeech and our in-house dataset, respectively.
The word/character error rate (WER/CER) were calculated for LibriSpeech and our in-house dataset, respectively. We used the ESPnet \cite{espnet} toolkit. 

\vspace*{-0mm}
\subsection{Effect of the joint training}
\label{sec:mainresults}
\vspace*{-1mm}
Table \ref{maintable} presents two forms of results, the effect of the joint training and decoding. First, we show that the single-decoder branches of our 4D models outperform their counterparts without joint training (\texttt{A1-4} vs. \texttt{B1-4}) -- 4D joint encoders improve CTC, attention, RNN-T, and Mask-CTC models by 0.28-0.72 WER/CER overall.
This improvement persists for CTC/attention models as well (\texttt{A5} vs. \texttt{B5}).
The effect of the joint decoding (\texttt{B6-8}) will be discussed in Section~\ref{sec:rtf-wer}

% Table \ref{maintable} shows the main results. Even without a joint decoding, the 4D model's attention decoder outperformed most of the baselines, confirming that the regularization effect of the multitask learning improved robustness. Since the computational cost of decoding is equivalent to the baselines unless joint decoding is used, the joint training of the 4D model improved performance without increasing the computational complexity. Furthermore, the results of joint decoding using two decoders (CTC/attention and RNN-T/attention) outperformed the results of a single decoder (attention) across three datasets. Thus, integrating multiple decoders during inference compensated for the shortcomings of each. Additionally, both CTC-driven and RNN-T-driven joint decoding with three decoders outperformed the joint decoding methods with two decoders. In other words, we reduced the disadvantages of each by incorporating three decoders. 
% For Librispeech 960h and CSJ, the performance of CTC/attention of the 4D model was comparable to that of the baseline, however it shows a significant improvement for Librispeech 100h. This method is especially effective with small amounts of training data as a regularization effect. 
% CTC-driven joint decoding performs slightly worse than RNN-T-driven joint decoding because of the pruning before RNN-T scoring, as described in Section \ref{sec:joint docoder}.
% Although, the decoder weights must be tuned in advance, the proposed method performs consistently across three datasets with the same decoder weights. 

\begin{figure}[t!]
\vspace*{-5.0mm}
     \centering
     \hfill
     \begin{subfigure}[b]{0.49\linewidth}
         \centering
         \includegraphics[scale=0.41]{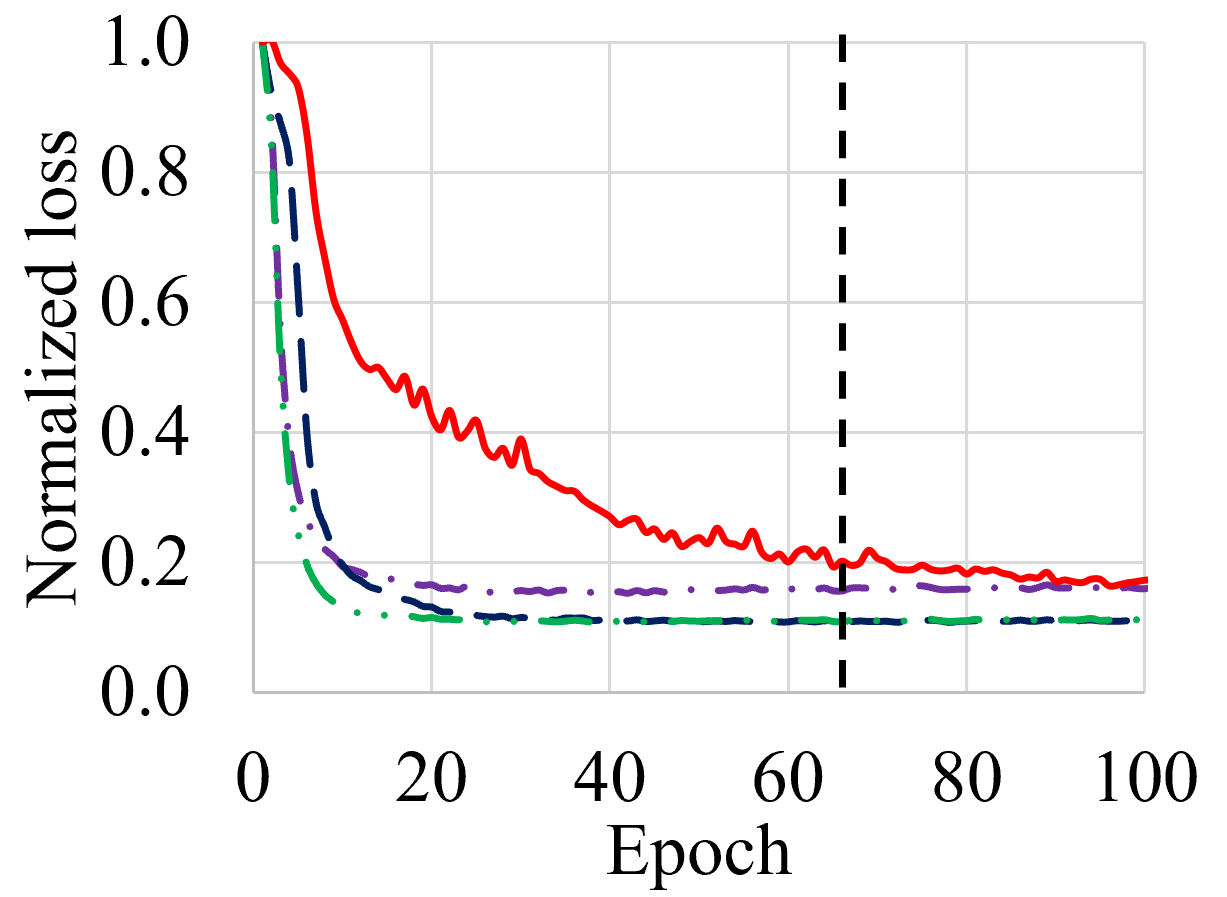}
         \vskip -0.07in
         \caption{First stage}
         \label{fig:stage1}
     \end{subfigure}
     \hfill
     \begin{subfigure}[b]{0.49\linewidth}
         \centering
         \includegraphics[scale=0.41]{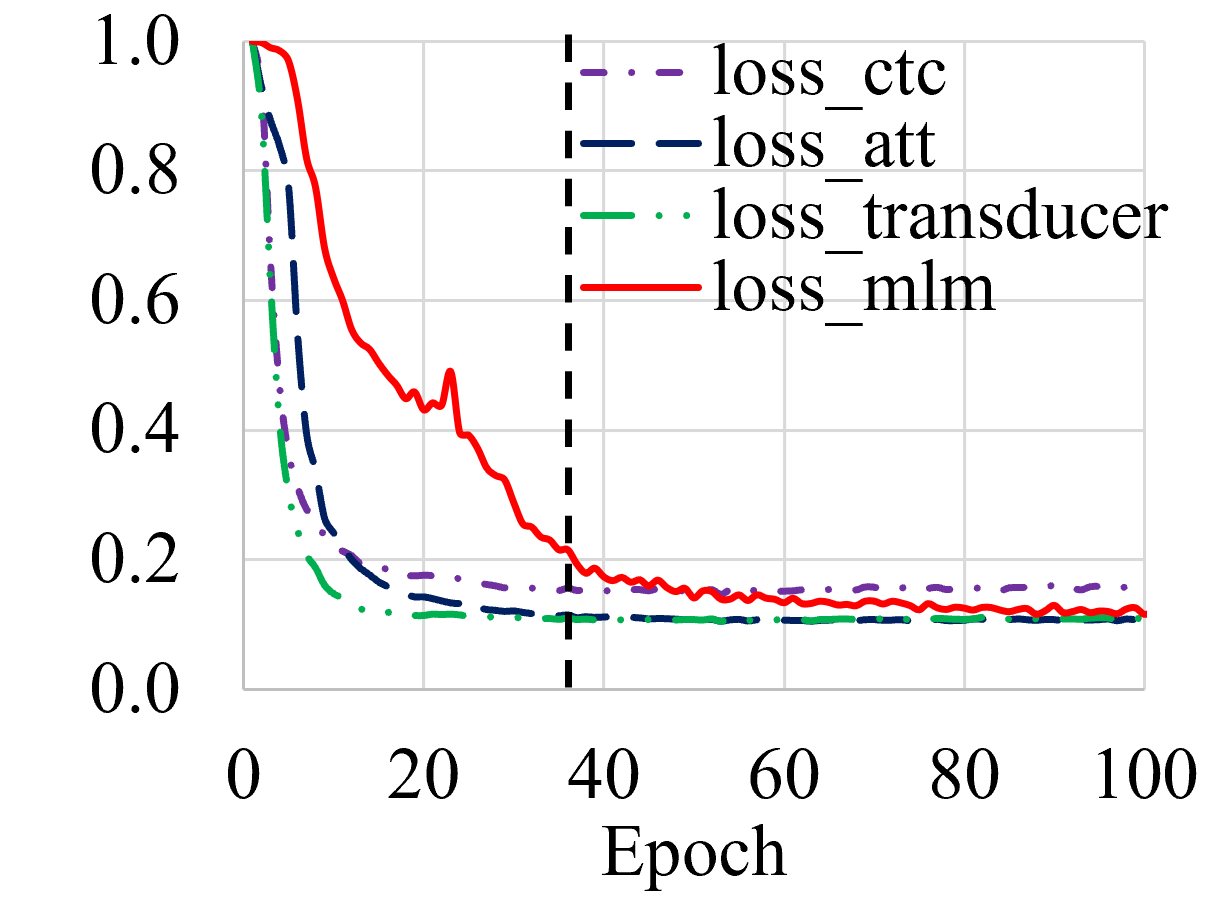}
         \vskip -0.07in
         \caption{Second stage}
         \label{fig:stage2}
     \end{subfigure}
     \hfill
    \vskip -0.12in
    \caption{Validation curves of the first vs. second stage.}
    \label{fig:stage12}
    \vskip -0.06in
\end{figure}

\begin{table}
\caption{Effect of the two stage strategy with Librispeech 100 h.}
\vspace{-7mm}
\label{weights}
\begin{center}
\begin{tabular}{@{}c|ccc}
\hline
Decoder & w/o multitasking & 1st stage & 2nd stage\\
\hline
Attention  & 19.43 & 18.69 & \textbf{17.91}\\
CTC        & 20.66 & 19.52 & \textbf{18.96}\\
Mask-CTC   & 20.78 & 20.38 & \textbf{18.97}\\
RNN-T      & 18.30 & 18.10 & \textbf{17.61}\\
\hline
\end{tabular}
\end{center}
\vspace*{-8mm}
\end{table}

\vspace*{-0mm}
\subsection{Analysis of the two-stage training strategy}
\label{sec:joint training}
\vspace*{-1mm}

Figure \ref{fig:stage12} shows the normalized validation losses in the first and second stages. %Note that the validation losses were normalized to compare the convergence speed rather than the magnitude of the losses. 
The MLM loss took more epochs to converge than the other three decoders in the first training (Figure \ref{fig:stage1}), indicating that the trained model did not converge sufficiently with the MLM decoder, or the other decoders were overfitted. 
The difference in the convergence speed of the four losses, on the other hand, was smaller in the second training, indicating that the four losses converged relatively adequately (Figure \ref{fig:stage2}).
Table \ref{weights} shows the performance of each decoder on the test-other set of Librispeech 100 h without and with joint training in the first/second stage.
Even the first stage outperformed the model without multitask learning, 
%However, Mask-CTC performed worse than the other three models. 
the performance of all four decoders improved in the second stage. 
%because all decoders converged without under/overfitting by tuning the training weights.
Using the proposed two-stage approach, the four weights were efficiently determined with only two experimental trials.

\begin{figure}[t!]
\vspace{-6mm}
    \centering
        \begin{minipage}{0.45\textwidth}
            \includegraphics[width=\textwidth]{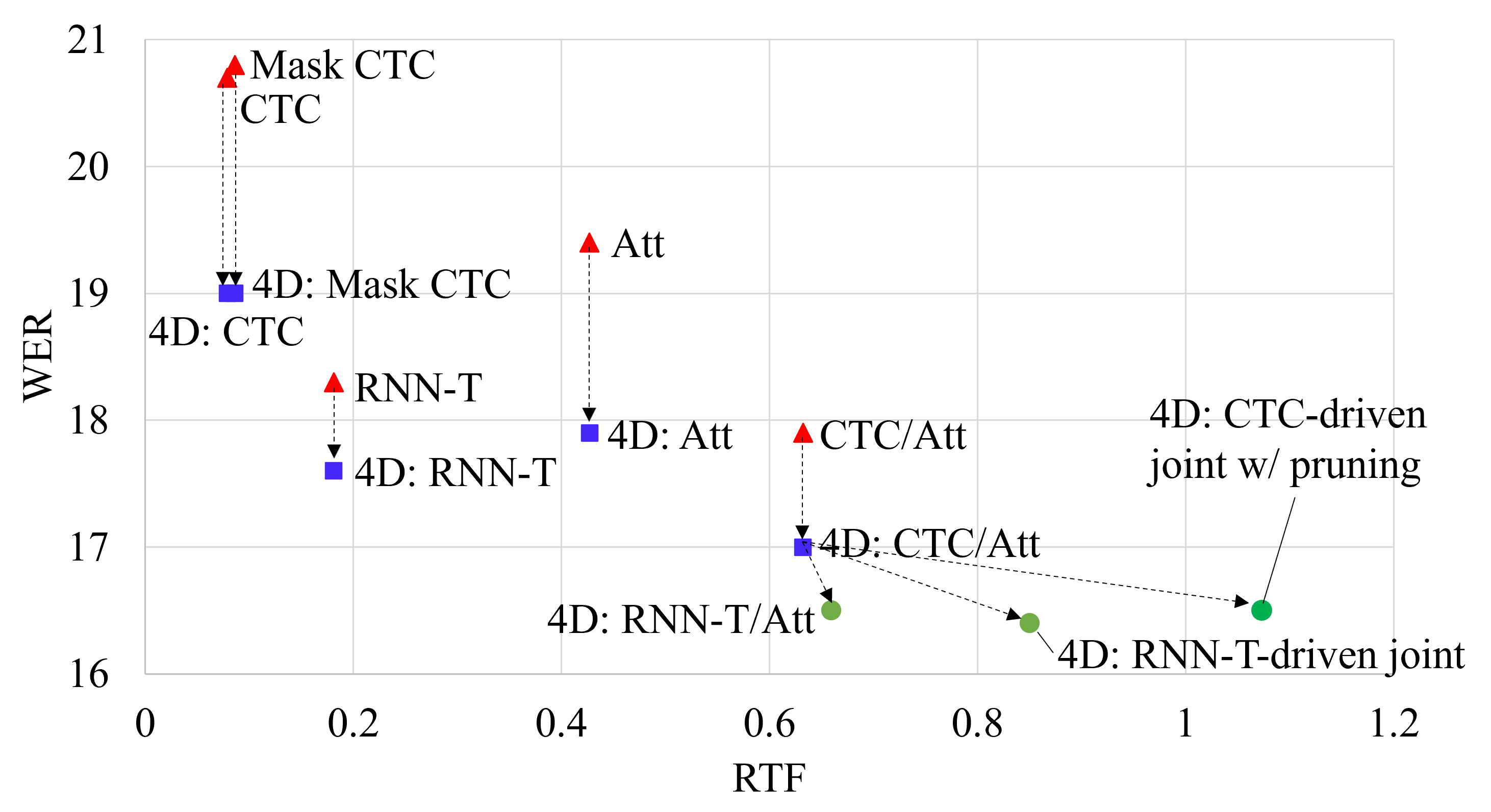} 
        \end{minipage}
        \centering
    \vspace*{-3.5mm}
    \caption{Relationship between RTFs and WERs. The red, blue, and green dots denote the baselines, 4D model without joint decoding, and 4D model with joint decoding, respectively.}
    \label{fig:rtf}
\vspace*{-3mm}
\end{figure}

\vspace*{-0mm}
\subsection{Effect of the joint decoding}
\label{sec:rtf-wer}
\vspace*{-1mm}

Table \ref{maintable} also presents the results of joint decoding (\texttt{A5} and \texttt{B6-8}).
We show that 4D models also offer RNN-T/attention and CTC/RNN-T/attention decoding, which outperform CTC/attention on average (\texttt{A5} vs. \texttt{B6-8}) -- RNN-T-driven CTC/RNN-T/attention decoding is 0.47 WER/CER better than CTC/attention overall.
Note that these improvements are more pronounced on Librispeech 100 h, suggesting that the 4D method offers a regularization effect which is important for smaller amounts of training data.

Figure \ref{fig:rtf} shows the relationship between real-time factor (RTF) using a GPU (NVIDIA RTX3090) and WER on the Librispeech 100 h test-other set. The red dots denote the baselines, the blue dots denote 4D joint training but \textit{without} joint decoding, and the green dots denote 4D joint training \textit{with} joint decoding. Comparing the red and blue dots, the proposed 4D model reduced WER for all decoders without increasing RTF 
%Although the joint training requires more GPU memory and training time, the proposed 4D model does not increase the computational cost as discussed in Section \ref{sec:mainresults}
as long as a single decoder is used. The proposed two joint decoding methods had larger RTFs than the other decoders due to increased complexity, but the WERs were the smallest. 
CTC-driven joint decoding had a larger RTF than RNN-T joint decoding, even with the pruning as described in Section~\ref{sec:joint docoder}.

\vspace*{-0mm}
\section{Conclusion}
\label{sec:conclusion}
\vspace*{-0mm}
This paper proposed a 4D joint model of CTC, attention, RNN-T, and Mask-CTC by sharing an encoder trained in a multitask fashion. 
We demonstrated that jointly trained 4D models with the proposed two-stage training strategy improved the performance of each individual decoder. 
Furthermore, the proposed joint CTC/RNN-T/attention decoding improved the performance and outperformed the previously proposed CTC/attention decoding.

%In the proposed 4D model, each single decoder can be easily switched for CTC, attention, RNN-T, and Mask-CTC depending on the application scenarios.
%Such a universal ASR model is practical because different application scenarios typically have different requirements.
%Additionally, the proposed two-stage training strategy enabled efficient tuning of the training weights and outperformed conventional E2E ASR models. 
\clearpage

\bibliographystyle{IEEEtran}
\bibliography{mybib}

\end{document}